**Assembling a Cyber Range to Evaluate Artificial Intelligence / Machine Learning (AI/ML) Security Tools**

Jeffrey A. Nichols, Kevin D. Spakes, Cory L. Watson, Robert A. Bridges
Oak Ridge National Laboratory, Oak Ridge, TN, USA
nicholsja2@ornl.gov
spakeskd@ornl.gov
watsoncl1@ornl.gov
bridgesra@ornl.gov

**Abstract:** In this case study, we describe the design and assembly of a cyber security testbed at Oak Ridge National Laboratory in Oak Ridge, TN, USA. The range is designed to provide agile reconfigurations to facilitate a wide variety of experiments for evaluations of cyber security tools—particularly those involving AI/ML. In particular, the testbed provides realistic test environments while permitting control and programmatic observations/data collection during the experiments. We have designed in the ability to repeat the evaluations, so additional tools can be evaluated and compared at a later time. The system is one that can be scaled up or down for experiment sizes. At the time of the conference we will have completed two full-scale, national, government challenges (experiments designed for head-to-head evaluation of competing tools for over $150K in prizes per competition) on this range. These challenges are evaluating the performance and operating costs for AI/ML-based cyber security tools for application into large, government-sized networks. These evaluations will be described as examples providing motivation and context for various design decisions and adaptations we have made. The first challenge measured end-point security tools against 100K file samples (benignware and malware) chosen across a range of file types. The second is an evaluation of network intrusion detection systems efficacy in identifying multi-step adversarial campaigns—involving reconnaissance, penetration and exploitations, lateral movement, etc—with varying levels of covertness in a high-volume (up to 10Gb/s), business network. The scale of each of these challenges requires automation systems to repeat, or simultaneously mirror identical the experiments for each ML tool under test. Providing an array of easy-to-difficult malicious activity for sussing out the true abilities of the AI/ML tools has been a particularly interesting and challenging aspect of designing and executing these challenge events. After the events, the range continues to be used for other research such as adversarial machine learning where the repeatability, scale, and automation required for the national challenge events become essential elements for research.

## 1. Description

This is a case study that describes our work to build a network-isolated, test environment where we can test and evaluate the performance and capabilities of cyber security systems including live user-studies involving these systems. The immediate predecessor to this range was a 25-node cluster we built using Emulab (White et al, 2002 and www.emulab.net). Many of the initial design decisions derived from this experience. We have termed our system CORR, standing for Cybersecurity Operations Research Range. We have completed two world-class challenges on CORR to evaluate the current capabilities of AI/ML cyber security products.

For the first challenge evaluating endpoint malware detectors, the team compiled an enormous corpus of over 34M malware and benign-ware samples from a wide variety of sources, which permitted the largest-scale and most realistic experiment of its kind. Specifically, the team tested each competitor's detector on a curated selection of 100K files, two orders of magnitude greater than any previous evaluation set. To increase fidelity, files in the evaluation set were sampled to match the real-world distribution of filetypes, which ultimately revealed enormous gaps in detectors' abilities. Through collaborations with researchers at Assured Information Security (AIS), 1K zeroday (never-before-seen) malware were created and added to the set to test generalizability, the main benefit of machine learning tools.

Implementing a repeatable experiment of this scale required orchestration software engineered to run many simultaneous, identical tests. To prevent malware infections from affecting results, per-file instantiations of virtual machines, detection tools (under test), and performance- and accuracy-monitoring capabilities were choreographed into a completely automated, time-synced, and reproducible experimental framework, that parallelized 2.5K simultaneous instances. This software engineering feat reduced the experimental time from years per tool to 15 hours, allowing results that fit the deadline.

While the scale of the experiment allowed for comprehensive and fine-grained analysis, a scoring framework that translates the many experimental observations to a real-world context was needed to make the results actionable. A custom mathematical model was designed to take the many experimental measurements as input, outputting a comparable, quantifiable cost by simulating of how the tool would perform in Naval operations. This comprehensive cost model was motivated and informed by the teams' operations research, specifically, touring Naval security operations and interviewing operators.

The second challenge evaluated AI/ML-based network intrusion detectors. Realistic testing of network detection AI/ML tools required the creation of a unified network environment where the benign and adversarial network devices generating the traffic were indistinguishable, so that the only difference was in the network traffic itself. Most AI/ML systems also require periods of training to learn what is normal behavior. Several other issues also arose when creating a test environment for careful evaluation of AI/ML systems which are briefly described in this paper.

This paper is in two parts. The first several sections go through the Goals, Design, and Systems Implementation (Sections 2,3,4). The last section (5) is on lessons learned applying the system to the two national challenges evaluating AI/ML cyber security systems.

## 2. Design Goals

This range is the culmination of a long development series, going through various platforms and assemblages which focused on different goals. An early one emulated MIPS processors so that we could boot and run Cisco firmware images for high-fidelity network security testing. Another was a range for running live malware in a bare-metal environment, isolated from the corporate networks, that could be reset in minutes to a known, fixed state and experiments conducted in an automated way; this only supported one test subject at a time. We built an Emulab cluster prior to CORR using about 30 older servers that had been excessed from other projects. We added to these three 48-port networking switches and quad-port network cards, so each server had one management network and five network ports for testing. Emulab's focus is on simulating large and complex network topologies populated with virtual and bare-metal Linux and Windows servers and workstations that can model a corporate environment.

The goals for CORR derived from these previous projects. Foremost, we wanted a network-isolated range for cyber security experiments. These experiments were not pre-defined, but we did have in mind simulations with 1000 computers and users, so virtualization of Windows and Linux was a primary requirement. In addition to virtual machines, we wanted the ability to integrate various applications running on bare metal, such as mail and database servers, so the chosen hardware servers needed to be designed to flexibly handle both types of workloads. Network isolation—especially from the corporate network—was a requirement for both us and our IT security team. Some ranges take network isolation to an extreme with no Internet access. We designed ours so that we could verifiably control Internet and corporate network access when experiments are being conducted, especially those involving malware. Having Internet access when setting up systems speeds up setting up the range. With this isolation, we are free to reconfigure the systems and networks as needed. Our intent was to run the new range as a large Emulab cluster. Therefore, we had the concept of a BOSS node, many 1Gbps and 10Gbps ports per system, a management network, and network switches that could be programmed by Emulab's BOSS node to accurately emulate multiple networks simultaneously. Another feature from Emulab for which we designed was the ability to repeat experiments exactly using stored templates and configurations. We originally designed that all of the people involved with the experiments, such as SOC operators and red team members, would be on-site. We have since added remote management and access.

In summary, our goals for CORR were:
1) Isolated from the corporate network, especially for research involving malware;
2) Ability to turn Internet access on and off;
3) Servers capable of virtualized and bare-metal applications;
4) Virtualization of 1000 or more Linux and Windows machines;
5) Complex and accurately emulated network interconnections between systems; and,
6) Record and repeat experiments.

## 3. Design Ideas

In this section we describe the conceptual ideas we used when designing CORR. The diagram in Figure 1 was what we used to initially describe the system to others. This diagram illustrates the central features of CORR: 1) the experiment site being tested and evaluated, termed InnerCORR; 2) the systems wrapped around the test environment that are used for directing, storing, and evaluating the experiment, termed OuterCORR; and 3) the isolation of the networks and systems from the corporate environment. With the Emulated SOC, it also illustrates the idea of using CORR to conduct user studies in the environment to evaluate the interaction of cyber security operators with their information systems, while being able to inject real events for them to detect and evaluate. Thus, we needed to be able to support large information systems in the InnerCORR and provide the information feeds required by these information systems. In addition, we needed to provide workstations like what would be found in a SOC for the operators to use for the experiments. Similar comments apply to the OuterCORR also, particularly to the Analytic Stack, in which we needed enough storage to completely record the experiments so they could be repeated and to record the measurements for analysis.

The User Workstations, Emulated Network, and Network Services are illustrated as simple stacks in Figure 1. We wanted to simulate networks with multiple subnets, multiple services running on different servers, and connected physical hardware devices. Emulab has a tool that takes these sorts of network designs and builds and deploys the environment. Emulab records these deployments so the exact experiment can be restored later. Further, Emulab programs the network switches hosting the physical servers so that the subnets correspond to virtual LANs (VLANs) so that network traffic generated in the different subnets does not interact except by routing in order to match the behavior of real networks. (Emulab's direct configuration of switches requires selecting switches from a limited list of supported models.) Also, because of Microsoft licensing changes, Windows systems are no longer directly supported on Emulab. Because we required Windows, we purchased VMWare Enterprise ESXi licenses for a quarter of the virtual hosts. We were advised to also purchase VMWare vCenter; this turned out to be a vital component for our first AI/ML challenge since it provides robust automation and scripting of virtual machines and networks. Our previous experience with Emulab guided us towards making design choices compatible with Emulab that were also capable of being used on their own.

The final item we note from Figure 1 is that there are people involved in our experiments. A bulk of the equipment for CORR is located, by necessity, in a data center. We have support facilities in a separate building for use by the research team, operators, and the vendors. The buildings are joined by dedicated fiber, so our traffic never transits the corporate network. Most of the systems comprising OuterCORR are in the separate building. Also, the physical network is designed so that vendor equipment being evaluated can be installed in the support facility and incorporated into CORR without requiring access to the data center.

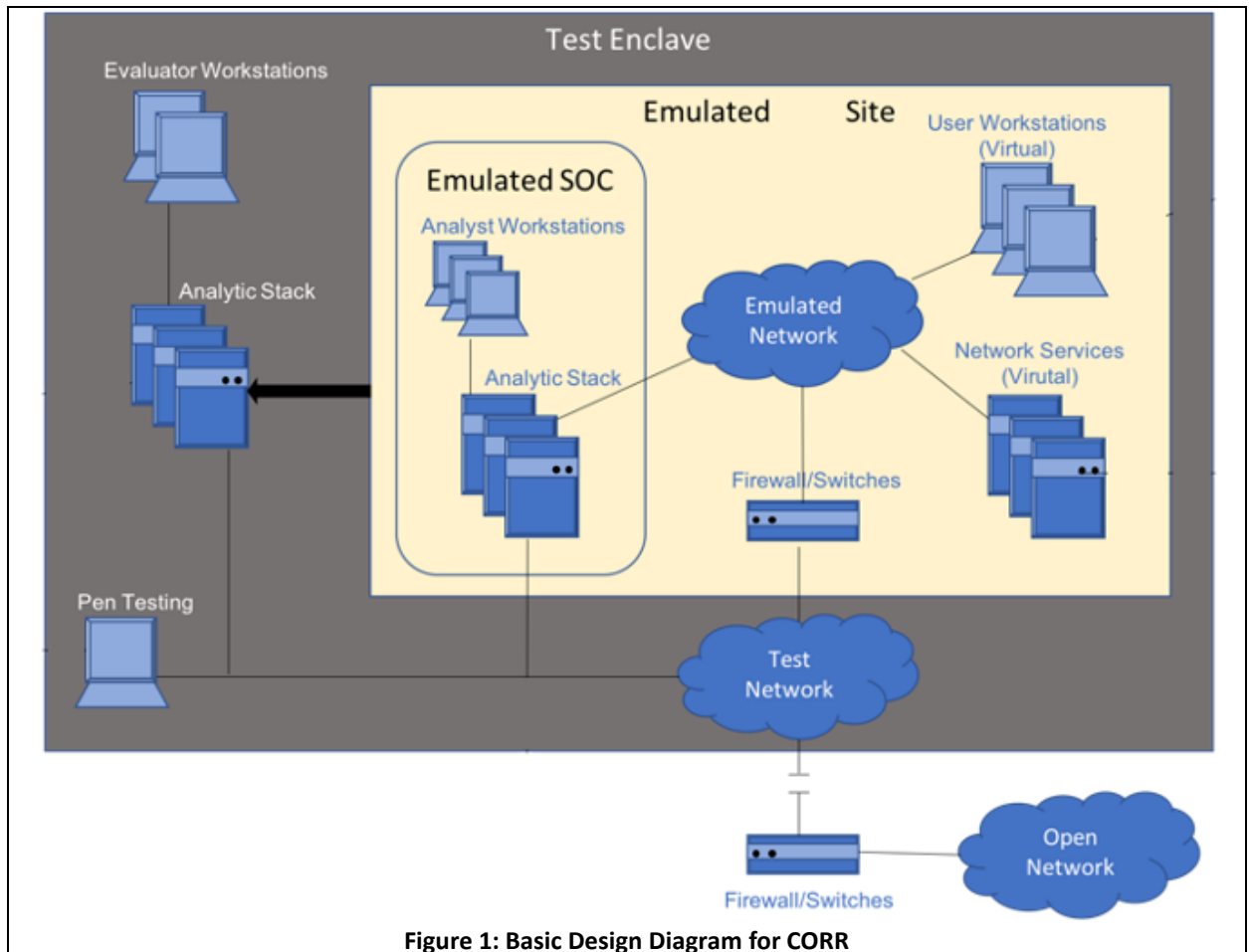

Figure 1: Basic Design Diagram for CORR

## 4. Component Systems

In this section we describe the major components of CORR. Additionally, we describe our reasoning and uses for particular hardware configuration choices we made.

### 4.1 Hardware

We now describe the hardware we chose for the worker nodes, the user systems, the experiment analysis, and the experiment storage.

#### 4.1.1 Worker Nodes

The worker nodes are the powerhouses of CORR. We needed the systems to perform as either virtualization hosts or bare-metal systems. Since bare-metal hosts would not typically require large core counts, we selected Dell PowerEdge R740xd servers because they are faster, 2-socket systems with a single drive controller. These can also perform adequately as virtualization hosts. For the systems intended as virtualization hosts, we chose the 4-socket, Dell PowerEdge R840 system with two drive controllers. For the network cards (NICs) we added quad-port, 1Gbit and 10Gbit network cards in addition to the built-in NICs. This was originally intended for Emulab but has proven to be valuable to physically separate different types of traffic. Large amounts of RAM, large numbers of cores, and dual drive controllers were chosen to increase the number of VMs each node could host. Often disk access is the constraint on VM performance. By splitting the disk arrays into two halves, then we are able to double the number of VMs the system can host.

Dell PowerEdge R740xd
- Processor:
  Intel Xeon Silver 4216 CPU @ 2.10GHz x2
- Logical Processors: 64
- Memory: 384 GB
- Storage: 52.3 TB
- NICs: 8

Dell PowerEdge R840
- Processor: Intel Xeon Gold 5117 CPU @ 2.00GHz x4
- Logical Processors: 112
- Memory: 384 GB
- Storage: 33.8 TB per drive controller
- NICs: 12

For the challenges we were able to easily host 50 Windows VMs and more Linux VMs per machine. Above this number we encountered performance issues. Because of the dual drive controllers, we were able to treat each R840 as two R740s. As a side note, VMWare is licensed by CPU socket, so the R740s are half the price to license.

### 4.1.2 User Systems

The user systems for the SOC analysts and penetration testing teams were standard, except for one powerful, GPU-enabled system for shared use. Since the red team members are cooperative members of our experiments, they can share resources, rather than requiring separate, powerful systems.

### 4.1.3 Experiment Analysis

The Experiment Analysis systems store data emitted by the systems being tested and store it in a streaming database like Elastic Search for analysis. Since it has similar requirements, it is composed of the same types of systems as the worker nodes as CORR. Because of this, the experiment analysis systems have ended up serving a second role as a miniature CORR; we use them to test our designs for upcoming experiments, while the main range is in use.

### 4.1.4 Experiment Storage

So that we can repeat experiments exactly, we require bulk storage of data. We wanted to use ZFS as the filesystem format, because of its data safeguards. The Dell Datacenter Scalable Server DSS7500 was recommended by our vendors. It is a 4U, data-vault design capable of holding 88 3.5-inch SAS drives. We populated about two thirds of the slots with 12TB SAS hard disks. In addition we added four 2.5" high-write endurance SSDs for ZFS caching. On this system we run FreeBSD because of its first-class support of ZFS. In conformance with ZFS best practices, we created five 11-drive pods which are unified by ZFS to present about 500TB of storage. We use iSCSI to export the drives to other systems because of its bandwidth.

## 4.2 Networking

We use four Dell 48-port S3148 (1Gbps) and three S4148 (10Gbps) switches to connect the components of CORR. Where we could, we used twinaxial cables for 10Gbit networking to keep costs down. The Dell switches are compatible with most of our 3rd-party SFP connectors (after disabling vendor checks), perform well, are full-featured, and are compatible with Emulab. Our two facilities are connected by fiber between two Dell S4148 switches. The network switches enable us to integrate virtual and bare-metal devices and add vendor devices to our CORR system for testing.

For our first challenge, we required little more from our network than fast connections between the virtual machines. For the second challenge the burden has been on the networking to produce the test environment. Another device we required for the second challenge is a packet splitter in order to present to the devices under test an identical packet stream. For this we use an Arista DCS-7150S switch in Tap Aggregation mode.

## 4.3 Software

### 4.3.1 Services

In Emulab designs, there is a single computer designated as BOSS that directs all of the simulations on the Emulab cluster. It also provides the web user interface for setting up experiments. As the only computer connected to both the outside network and the internal cluster, it provides the path through which users connect to their simulations and experiments. Any other outside connections and services required by the cluster computers also pass through or are provided by BOSS. We have maintained this design and use a single computer to provide: the Internet on/off switch and firewall as well as capturing connections made by software under test; services such as NTP, DNS, and DHCP; and remote access by our users.

VMWare vCenter Server is a virtual machine purchased from VMWare that runs on a VMWare ESXi cluster. ESXi machines can be managed without it and we did not think we needed this expensive, enterprise-grade virtual machine, but discovered that it is the vital component for performing VM automation. Because of its central role, we dedicate a single Dell R840 server to it. VMWare licenses are sold as either enterprise or standard. The enterprise features include automatic failover, mirroring, migration, and distributed virtual switches. We initially purchased enterprise licenses, but ended up turning these features off, because they interfered with our simulations. The only feature we have made use of is distributed networks, but its function can be duplicated, so we now use the less-expensive standard licenses.

*4.3.2 Operating Systems*

We use various versions of Linux on most of the systems. We use FreeBSD to manage our data vault, because of its native support for ZFS. Before settling on VMWare we also evaluated oVirt (Open Source Version of Red Hat Virtualization), XCP-ng (Open Source Version of XenServer), and KVM. Especially for Windows VMs, we struggled with these to get the performance we required. For us, the reliability and the automation tools for VMWare were worth the extra cost.

For the Windows virtual machines, since our use is for testing and we did not require any end-user features, we used the LTSC version of Windows from Microsoft. To achieve our startup time and testing goals we modified these machines to turn off unnecessary services. In addition, we disabled Windows Defender anti-virus and firewall, enabled NTP time sync, enabled network discovery, and installed Python 3.6.

## 5. Applying CORR to AI/ML Challenges

We have applied these components of CORR to two, national challenges—called Artificial Intelligent Applications to Autonomous Cybersecurity (AI ATAC)—evaluating AI/ML cyber security devices for their ability to correctly discern benign software and malware on the endpoint (DoD, 2019) and for their ability to detect attempted intrusions and other malicious actions using a full packet tap of a 1500-user corporate network (DoD, 2020). In this section, we describe how we applied CORR to these two challenges and we describe the lessons we learned carrying out these challenges.

### 5.1 Description

When designing these experiments, our effort was on making certain there was no inadvertent signal that would indicate if a sample was benign or malicious. In other words, we designed the experiment so that nothing stuck out that the AI/ML engines could easily learn and then base its decisions on. The steps we had to take to accomplish this was different for each challenge, so we describe the two challenges in separate sections.

### 5.2 AI ATAC 1 (Endpoint Protection)

*5.2.1 Experiment Description*

This first challenge called for AI/ML tools for detecting malware on endpoint systems. We received over a dozen submissions for evaluation. Most were software-only submissions, but several were combinations of hardware and endpoint software, referred to below as "devices." The rules of the challenge prevent the release of any specific information about the submissions. We requested and tested both Windows and Linux versions of the tools.

The challenge was wrapped around the simple experiment of presenting to the test device a known benign or known malicious sample file and recording its determination of whether or not the file was malicious. (Note that it is the device that is being tested, not the malware itself, so malware testing systems like Cuckoo are not applicable.) Also recorded was how long it took to make its determination, how many resources (CPU, memory, etc.) were used during its deliberation, and at what stage it made its determination. In the first stage, the sample was first presented statically to the device by placing it on its filesystem, so the device had the opportunity to perform static analysis on the sample; thus, matching known malware signatures was a valid detection strategy. We included samples to which the devices were naïve, so matching signatures was not a complete strategy. Most of the submissions included static analysis, often using several engines, as a component for making a decision. After approximately 92s, if a determination was not made by the device, the sample was loaded and executed and another minute was given for the device to make a determination. During this second stage, the sample's behavior and structure could be directly observed by the device and a determination made with the new information. Some of the devices would block execution, others sent out warnings, and others quarantined the files. These were all counted as making a determination that the file was malicious and the time to determination was recorded.

This simple experiment was repeated for 100,000 samples for each device and for the Windows and Linux versions of the devices. Half of the samples were benign. These samples were drawn from a previous summer student project, that scoured the Internet collecting millions of benign samples. These were classified by types: ISOs, Word documents, Windows executable, Java programs, etc., and our selections were made to maintain the distribution of types typical on business systems and on the Internet. The malicious samples were likewise drawn from an archive of over 30 million files we were given from virusshare.com. The malicious samples were also classified by the same types as the benign files and the distribution of these different types was maintained in our sample selections. After the experiments were completed, then true and false positive and negative rates were calculated and the devices were scored using a cost model (Iannacone and Bridges 2020) that included the time it took to reach a determination, the cost of the device, the cost of the resources used including labor, the incident response costs of false positives and false negatives, and the costs saved by detecting true positives.

The rules of the challenge required that the tools all be pre-trained and had to be hosted completely on-site. In other words, they could not send out samples for analysis nor receive determinations from an outside source. The devices could not be updated nor communicate out after the submission date to preserve our zero-day samples. We enforced these conditions by blocking any Internet access for the devices using the Boss node.

### 5.2.2 Experiment Execution

When building a system on which to execute this challenge, we went through several different hardware and software configurations. We report here on the final configuration we used. We used VMWare ESXi on eight R740 and eight R840 PowerEdge servers. We used the trial version of ESXi because we needed the systems for fewer than 60 days. We used vCenter on another ESXi host and developed python scripts that ran on vCenter to automate the testing and data collection. This framework coordinated about 2000 simultaneous VMs active across the cluster. Each VM ran for less than five minutes. We used another machine as a web server providing the 100K samples over http. Splunk was run on a different server to collect the results. As a result, the time required to execute the challenge was reduced from a year per tool down to about 16 hours per tool with all of the range operations required to stand-up each tool. We completed the entire suite in about three weeks.

For each tool and operating system version we created a template image in which the tool was installed. Windows Defender was disabled, except when we were testing it specifically, so that it did not interfere with the tests. This template image was first replicated once to each node in the cluster, then these were replicated on each node until the full number of machines was created. We treated each R840 system with two drive controllers as two separate R740s, so that the drive images went on separate controllers and CPUs. The lifecycle of a VM was: one minute to boot; one minute to download and statically examine the assigned sample; one minute to execute the sample; and a minute to complete data collection and notify the framework. Each tool being tested was configured to directly send its detection results to Splunk. After these steps were completed, then another VMWare feature, in-memory snapshots, reset the machine back to its initial state. Using the in-memory snapshot was much faster than reloading the VM cleanly from disk each time, because of disk controller contention.

All of the simulation under the control of a framework we developed in python. The scripts made extensive use of the VMWare API through the open-source python library, pyVmomi. Initially we interacted directly with the VMWare API. We found the API to be extensive and to cover every feature we required, but the documentation was lacking examples. Examples from the pyVmomi community were invaluable. Earlier iterations of our tests used APIs from other vendors, such as Virtualbox, but these were not as complete nor as robust as the VMWare API. Apache Kafka was used to broadcast the current state of the challenge, so it could be monitored by the framework. The framework kept track of tests and repeated any tests that were incomplete. We did have some samples that caused repeated crashes with some tools, so we had to establish a maximum number of attempts for each sample.

Between each tool the range had to be cleaned up by deleting the old images and distributing the new ones. The first version of this script took about eight hours to complete. We discovered from this script that the vmWare API has a maximum limit of 2,000 API connections at a time and it is limited to 600 concurrent tasks before queueing. Exceeding these values caused the scripts or image deployments to fail. We were not able to get around these hard-limits, but later we were able to reduce the redeployment to less than an hour by using a fast clone option.

Because resetting the range was relatively slow, we added a Quality Assurance/Quality Control (QA/QC) step at the end of the runs before resetting the range to confirm that all of the testing was complete. We also had a short QA/QC run before the full test of about 1500 representative samples to catch misconfigurations or software failures.

The most problematic network service we had was with synchronizing time on the machines through NTP. Windows machines prefer to use Internet time sources rather than NTP, but there was no Internet access. Windows also does not normally adjust its clock in the first minute of startup. We also found that, because we were dragging down the performance of the Virtual Machine Hosts, time measurement also dragged. We put a lot of effort into coordinating time to make the scoring results easiest to interpret, but we still did get time shifts on some machines for which we had to correct when scoring.

### 5.3 AI ATAC 2 (Network Detection)

#### 5.3.1 Experiment Description

The second AI/ML challenge called for network intrusion detection tools. We received about a dozen tools for evaluation; most are hardware systems with a few software-only systems.

The simple experiment for this challenge is to actively carry out attacks against vulnerable machines and to present the resulting network packet stream to the devices being evaluated. We measure the ability for each of the devices to discern the time and types of attacks. In contrast to the first challenge, all the devices are tested at the same time and are presented identical packet streams. To simulate the presence of 1000+ users, we combine in actual background traffic at the average rate of 1.25Gbps. Some tests evaluate if the tools' detection capabilities are affected when the rate of background traffic is quadrupled. We also vary the subtlety of the attacks and include attacks to which the systems are naïve to proxy for zero-days. See Figure 2.

#### 5.3.2 Experiment Execution

There are many more facets to executing this challenge, which are complicated by needing to prevent there being any easy clue for the AI/ML devices to pick up on. A source of simplification is that this challenge only involves what traverses the network. We present a single, full network tap to each device; there are no other sources of information like log files or Active Directory information. Because they utilize machine learning, most of the tools required a training period of several weeks to learn what is normal in the environment so that they could detect anomalous behavior. During this training period and testing, any machines we introduced needed to behave indistinguishably from the background systems. The different facets of executing this challenge are described in three groupings: Red Team, User-Emulation, and Experiment Infrastructure.

Red Team: We wanted to evaluate the devices using both professional pentesters and automated intrusion tools. We evaluated several tools to automate attacks and discovered Scythe (scythe.io) to be the most useful to us for automating practical attack methods. We also used pyMetasploit libraries to execute basic attacks. Some portions of the AI ATAC 1 framework were also used to transfer malware across the network and evaluate if the devices detected malware attempting to establish command-and-control. All of the network traffic is recorded.

User Emulation: After assembling the list of attack vectors, we created machines that would be vulnerable to these attacks. We also created other machines under our control for reacting to reconnaissance and probes. In all we created about 400 machines for the simulation which interact with each other and the network and report when they are probed. These machines had to behave like other machines in the background traffic, so we developed a data-driven, user-emulation system that creates online traffic including login/logout events. Additionally we created ways for the machines to exchange mail, visit common websites, interact with servers, and other common behaviors to prevent these machines from sticking out. These machines were assigned IP addresses spread throughout the range of addresses in use in the background traffic.

Experiment Infrastructure: All of this automated user-emulation and recording of measurements for device analysis requires additional infrastructure to execute without including these data in the packet stream presented to the devices. This was carried out by attaching additional NICs to our test machines which are used for recording measurements and for executing the user-emulation. This keeps the network data on the test network clear of extraneous traffic. Each of the tested devices is required to monitor 10Gbit networks. They each had a separate, 1Gbit management interface on which it reports its determinations and presents the UI for configuration. In addition, several devices had multiple components, so this management network also may be used for communications between the components. We used VLANs to isolate the traffic from other devices and additional, BOSS-type machines to provide remote access to each device's management UI. NTP time synchronization again plays a critical role for scoring the devices accurately. Separate Splunk instances were created for each device to report its findings. This prevents a device from gaming the system by preventing or delaying access for other competitors to the scoring systems.

We discovered that VMWare will optimize internal host network traffic, so we designed all communicating machines to transit the network switch so they are captured.

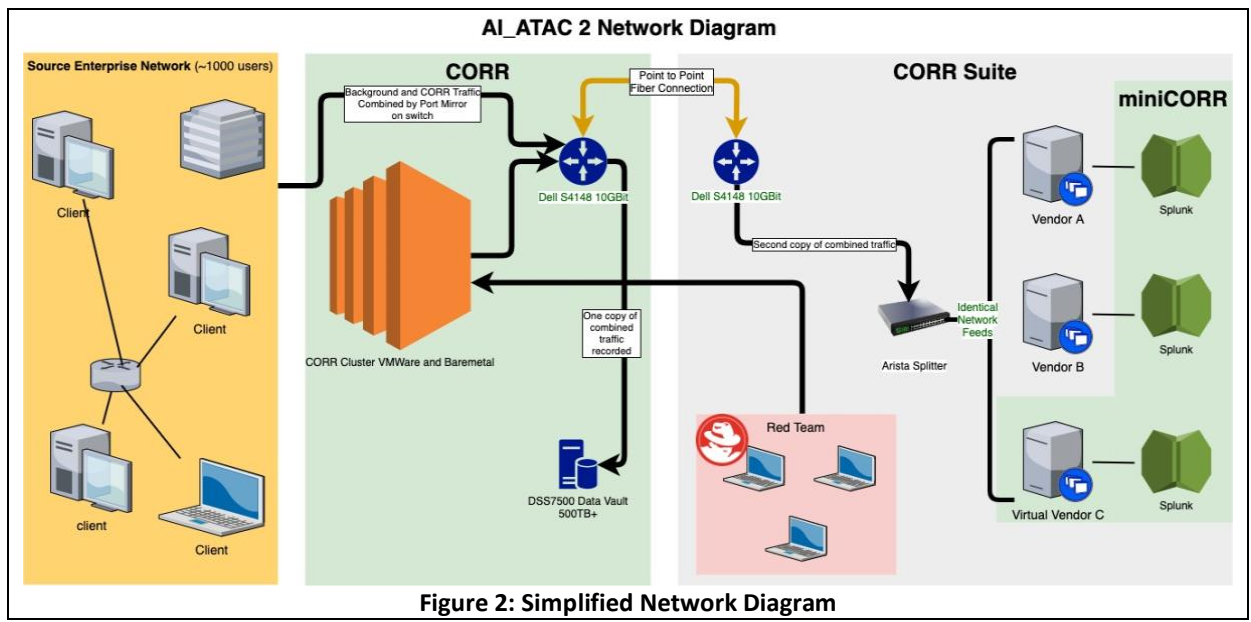

Figure 2: Simplified Network Diagram

## 6. Conclusions

CORR has mostly been used for executing national, cyber security challenges. It has also been useful to us for related research like Adversarial Machine Learning using the knowledge we gain by conducting the challenges. The equipment and design chosen has proven flexible to meeting the large-scale testing and evaluation challenges we have for it.

## Acknowledgements

This manuscript has been authored by UT-Battelle, LLC, under Contract No. DE-AC05-00OR22725 with the U.S. Department of Energy, and is based upon work supported by the Department of Defense (DOD), Naval Information Warfare Systems Command (NAVWAR). The views and conclusions contained herein are those of the authors and should not be interpreted as representing the official policies or endorsements, either expressed or implied, of the DOD, NAVWAR, or the U.S. Government. The United States Government retains and the publisher, by accepting the article for publication, acknowledges that the United States Government retains a non-exclusive, paid-up, irrevocable, world-wide license to publish or reproduce the published form of this manuscript, or allow others to do so, for United States Government purposes. The Department of Energy will provide public access to these results of federally sponsored research in accordance with the DOE Public Access Plan (http://energy.gov/downloads/doe-public-access-plan).